# Structure and Giant Magnetoresistance of Electrodeposited Co/Cu Multilayers Prepared by Two-Pulse (G/P) and Three-Pulse (G/P/G) Plating

N. Rajasekaran[a,b], L. Pogány[a], Á. Révész[c], B.G. Tóth[a], S. Mohan[b], L. Péter[a], I. Bakonyi[a,*]

[a]*Wigner Research Centre for Physics, Hungarian Academy of Sciences.*
*H-1121 Budapest, Konkoly-Thege út 29-33, Hungary*

[b]*Central Electrochemical Research Institute.*
*Karaikudi-630006, Tamil Nadu, India*

[c]*Department of Materials Physics, Eötvös University.*
*H-1117 Budapest, Pázmány Péter sétány 1/A, Hungary*

**Abstract** – The giant magnetoresistance (GMR) was investigated for electrodeposited Co/Cu multilayers. In order to better understand the formation of individual layers and their influence on GMR, multilayers produced by two different deposition strategies were compared. One series of Co(2 nm)/Cu($t_{Cu}$) multilayers with $t_{Cu}$ ranging from 0.5 nm to 6 nm was produced with the conventional two-pulse plating by using a galvanostatic/potentiostatic (G/P) pulse combination for the magnetic/non-magnetic layer deposition, respectively, whereby the Cu layer deposition was carried out at the electrochemically optimized potential. Another Co(2 nm)/Cu($t_{Cu}$) multilayer series with the same $t_{Cu}$ range was prepared with the help of a G/P/G pulse combination. In this latter case, first a bilayer of Co(2 nm)/Cu(6 nm) was deposited in each cycle as in the G/P mode after which a third G pulse was applied with a small anodic current to dissolve part of the 6 nm thick Cu layer in order to ensure the targeted $t_{Cu}$ value. The comparison of the two series revealed that the G/P/G pulse combination yields multilayers for which GMR can be obtained even at such low nominal Cu layer thicknesses where G/P multilayers already exhibit bulk-like anisotropic magnetoresistance only. Surface roughness measurements by atomic force microscopy revealed that the two kinds of pulse combination yield different surface roughness values which correlate with the structural quality of the multilayers as indicated by the absence or presence of multilayer satellite reflections in the X-ray diffraction patterns. A separation of the superparamagnetic (SPM) contribution from the total observed GMR provided useful hints at the understanding of differences in layer formation between samples prepared with the two kinds of pulse combination. The results of multilayer chemical analysis revealed that mainly an increased Cu content of the magnetic layer is responsible for the onset of SPM regions in the form of Co segregations in the G/P/G multilayers with small Cu layer thicknesses. Magnetization measurements provided coercive force and remanence data which gave further support for the above interpretation of the GMR data.

---

*Corresponding author. E-mail: bakonyi.imre@wigner.mta.hu

# Introduction

The giant magnetoresistance (GMR) effect in electrodeposited (ED) multilayer films was extensively studied in the last two decades.[1] Although several magnetic/non–magnetic element combinations are accessible for the electrodeposition process to produce multilayers exhibiting a GMR effect, particular attention has been paid to the Co–Cu system (see detailed list of references in Ref. 1). This is mainly due to the fact that Co/Cu multilayers prepared by physical methods were found to exhibit the largest GMR effect[2,3] with an oscillatory behavior of the GMR magnitude as a function of the Cu spacer layer thickness.

As to the GMR in ED Co/Cu multilayers, various dependencies of the GMR magnitude on spacer thickness have been reported.[1] It has been pointed out, however, in a recent paper[4] that the observed spacer thickness dependence of GMR in ED multilayers can be strongly influenced by the presence of a superparamagnetic (SPM) contribution to the GMR (Ref. 5), especially at low spacer thicknesses. On the other hand, the oscillatory behavior of GMR commonly observed in physically deposited multilayers[2,3] derives from an oscillatory exchange coupling between adjacent ferromagnetic (FM) layers. Therefore, when ED and physically deposited Co/Cu multilayers are compared, the FM contribution to the GMR (Ref. 4) should only be displayed as a function of the Cu layer thickness.

Confining to reports on ED Co/Cu multilayers in which the FM contribution to the GMR could be unambiguously identified, it turns out[4] that for low Cu layer thicknesses (typically below 1 to 2 nm) an anisotropic magnetoresistance (AMR) effect characteristic of bulk ferromagnets is only observed with a small or negligible GMR contribution. This can be ascribed to a direct FM coupling between adjacent magnetic layers through pinholes in the Cu spacer in this layer thickness regime. This is strongly supported by the coercive field ($H_c$) values[4] as low as 20 Oe, typical for bulk ferromagnets. For higher Cu layer thicknesses, the GMR increases monotonically and reaches a maximum in the Cu thickness range of about 3 to 5 nm after which the GMR decreases. The presence of pinholes at low spacer thicknesses was demonstrated experimentally by cross-sectional TEM for sputtered Co/Cu multilayers by Bobo et al.[6] and these authors have also studied the influence of pinholes on the interlayer coupling by comparing measured and modeled magnetization curves of the multilayers.

It was also concluded in Ref. 4 that the monotonous increase of GMR in ED Co/Cu multilayers with Cu layer thickness arises due to the diminished FM coupling between adjacent magnetic layers as the Cu layers separate more and more perfectly (and, hence, magnetically decouple) the magnetic Co layers. This facilitates, on the other hand, a more and more random distribution which, yielding at the same time a lesser and lesser aligned state of the individual layer magnetizations in zero field, is an important prerequisite for an increased GMR. The concomitant increase of the $H_c$ values with Cu layer thickness toward a saturation of the coercive field characteristic for individual, non-interacting magnetic layers provided a further support for this picture. Whereas it was not possible to reveal the very origin of the lack of an antiferromagnetic (AF) coupling at any Cu layer thickness, it still became clear[4] that a key issue toward improving the size of the GMR effect in ED Co/Cu multilayers is the reduction of the critical Cu layer thickness above which the adjacent magnetic layers are in an essentially decoupled state. The major contribution to the FM coupling at low Cu layer thicknesses is certainly via a direct contact through pinholes in the Cu layers although the so-called "orange-peel" coupling[7,8] due to layer undulations may also play a role.

It was, therefore, decided to try alternative routes for controlling the Cu layer continuity. Along this line, the aim of the present work was to compare the Cu layers in ED Co(2 nm)/Cu($t_{Cu}$) multilayers (0.5 nm ≤ $t_{Cu}$ ≤ 6.0 nm) with the formation of a spacer layer by the conventional two-pulse plating via a galvanostatic/potentiostatic (G/P) pulse combination and by a novel three-step process with the help of a G/P/G pulse combination. When applying the G/P pulse sequence, a Co/Cu bilayer is formed in each cycle by depositing first a Co layer by galvanostatic deposition which is followed by a potentiostatic deposition of the Cu layer. With the G/P/G sequence, a Co(2 nm)/Cu(6 nm) bilayer is deposited first in each cycle as in the G/P mode after which a third G pulse is applied with a small anodic current to dissolve part of the 6 nm thick Cu layer in order to achieve the targeted $t_{Cu}$ value. It should be noted at this point that the Cu dissolution pulse leads to an increase of the local $Cu^{2+}$ ion concentration. As a consequence, the next pulse for the magnetic layer deposition will result in an increased Cu content in this layer which even changes with thickness. Therefore, the so-called magnetic layer for the G/P/G multilayer is not well defined and even its effective thickness may vary depending on the degree of Cu layer dissolution.

For the P pulse of both pulse sequences, the deposition potential of the Cu layer was optimized so as to avoid both the dissolution of the previously deposited Co layer and the codeposition of Co in the Cu layer. At the optimized potential, the Cu deposition takes place at the diffusion-limited current density. In this manner, it is ensured that the actual layer thicknesses will correspond well to the preset nominal values; of course, with the uncertainty of the magnetic layer thickness for the G/P/G series as discussed in the previous paragraph.

The comparison of the two series revealed that the G/P/G pulse combination yields multilayers for which GMR can be obtained even at such low nominal Cu layer thicknesses where G/P multilayers already exhibit bulk-like anisotropic magnetoresistance only. However, the G/P/G pulse combination also resulted in the formation of a large amount of SPM regions in the multilayers and this necessitated the decomposition[5] of the SPM contribution to the measured GMR.

In addition to measuring the magnetoresistance, magnetic hysteresis loops were also recorded in order to extract some information on the evolution of the coercive filed and the remanence of the multilayers in both series.

Furthermore, an overall chemical composition analysis as well as X-ray diffraction (XRD) and atomic force microscopy (AFM) measurements have been carried out in order to characterize the structure and the surface roughness of the multilayer samples.

**Experimental**

*ED Co/Cu multilayer preparation and characterization.* — The magnetic/non-magnetic Co/Cu multilayers were prepared from an aqueous electrolyte containing 1 M $CoSO_4$ and 0.025 M $CuSO_4$. The multilayer electrodeposition was performed on a Si(100)/Cr(5 nm)/Cu(20 nm) substrate where the Cr adhesive and Cu seed layers were obtained by evaporation. Electrodeposition was carried out in a tubular cell[9,10] at room temperature in which the substrate was at the bottom of the cell with upward looking cathode surface area of about 7.5 mm by 20 mm. This arrangement ensures a lateral homogeneity of the deposits and helps to avoid edge effects.

For the present study, Co/Cu multilayers were electrodeposited with a constant value of $d_{Co}$ = 2.0 nm for the magnetic layer thickness and with the Cu spacer layer thickness $t_{Cu}$ varying from 0.5 nm to 6 nm. The bilayer number was varied in a manner to maintain a total multilayer thickness of 300 nm.

Two series of samples were produced. In the first case (G/P series), the conventional two-pulse plating was applied in the mixed galvanostatic/potentiostatic (G/P) deposition mode[9] in which the magnetic layer (a Co-rich Co-Cu alloy) is deposited by controlling the deposition current (G mode), whereas the non-magnetic layer (pure Cu) is deposited by controlling the deposition potential (P mode). The magnetic layer deposition was carried out at fixed cathodic current density amplitude of -50 mA/cm$^2$. According to a detailed analysis on a large set of multilayers prepared under similar conditions,[11] for such multilayers the current efficiency during the magnetic layer deposition is almost unity. Therefore, the nominal magnetic layer thicknesses were calculated on this basis from Faraday's law.

For the Cu layer deposition, the deposition potential was optimized.[12] For this purpose, a cyclic voltammetric study was first performed with the electrolyte used and the optimization was then performed by using a chronoamperometric technique. A copper foil was used as a counter electrode and the potential was referred to a saturated calomel electrode (SCE). Figure 1a displays the cyclic voltammogram of the electrolyte used for the deposition of Co/Cu multilayers by using a Pt substrate. Here, the copper reduction takes place in the range of -250 mV to -800 mV (see inset in Fig. 1a) and then, the current flow is increased due to the onset of cobalt reduction. The optimized copper deposition potential is determined during two-pulse plating experiments through a chronoamperometric technique by recording the current transients at several Cu deposition cathode potentials immediately following the G pulse for magnetic layer deposition and the results are shown in Fig. 1b.

The current transients are studied during the copper deposition pulses. To this end, the current density variations are monitored systematically in the appropriate potential range revealed from cyclic voltammetry for a better understanding of the occurrence of anodic and cathodic transients. When the cathode potential is sufficiently positive with respect to an optimum Cu deposition potential, there is a positive current flow which is an indication of the dissolution of the already deposited cobalt layer. When the potential is driven towards more negative values, the positive dissolution current reduces and we can observe the rapid onset of a steady-state current at -600 mV. Beyond this potential, the current flow is higher than the steady state current and this reveals the cobalt codeposition into the copper layer. The fastest steady-state transient is reached at the optimum potential (in this case -600 mV) but even here, we can still observe a very small initial positive current flow due to the capacitive nature of the electrode/electrolyte interface. The steady-state diffusion-limited Cu deposition current density was -1.4 mA/cm$^2$. The Cu layer thickness was set by measuring the charge passed through the cell and by using Faraday's law under the usual assumption of 100 % current efficiency for Cu deposition at the limiting current density. From the ratio of the diffusion-limited Cu deposition current density (P pulse) to the current density used for Co-layer deposition (G pulse), it can be estimated that the Cu content of the magnetic layer is about 2.8 at.% when producing the Co/Cu multilayers with the G/P pulse sequence.

The second series of Co/Cu multilayers was grown with the help of a G/P/G pulse combination. After depositing first a magnetic Co(2 nm) layer with a G pulse in an identical manner as for the G/P series, a Cu layer of 6.0 nm thickness was grown potentiostatically (P pulse) at the above described optimized potential. The last G pulse in which an anodic current density amounting to about 1/5 of the limiting Cu deposition current density in the bath used served for the fine-tuning of the Cu layer thickness by gradually dissolving the 6 nm thick Cu layer deposited in the P pulse. The layer thickness change during dissolution was also calculated from Faraday's law by assuming $Cu^{2+}$ formation with 100 % current efficiency.

The overall composition of the multilayers was determined by electron microprobe analysis in a JEOL JSM840 scanning electron microscope equipped with a RÖNTEC analytical facility.

Structural information was obtained by using XRD and the lattice parameters were calculated by using a crystallographic least-square refinement process. Lorentzian curves were fitted to the background-corrected XRD diffraction patterns to determine the peak positions and the full width at half-maximum (FWHM) value of the peaks. Where multilayer satellite reflections were observed, the bilayer thickness was calculated from the satellite peak positions.

The root-mean-square surface roughness ($R$q) of the deposited multilayers was determined by using atomic force microscopy with an Agilent Technologies 5500 instrument.

*Magnetoresistance and magnetic measurements.* — The magnetoresistance measurements were performed at room temperature with the four-point-in-line method in magnetic fields $H$ between –8 kOe and +8 kOe in the field-in-plane/current-in-plane geometry. Both the longitudinal (LMR) and the transverse (TMR) magnetoresistance (field parallel to current and field perpendicular to current, respectively) components were recorded for each sample. The following formula was used for calculating the magnetoresistance ratio: $\Delta R/R_o = [R(H) - R_o]/R_o$ where $R(H)$ is the resistance in the magnetic field $H$ and $R_o$ is the resistance maximum value around $H = 0$. A shunting-effect correction due to the metallic underlayers on the substrate was done on the measured MR data by using the measured values of the zero-field resistivity of both the substrate and the substrate/multilayer stack.[13] The observed $MR(H)$ curves were decomposed according to a standard procedure[5] into FM and SPM contributions of the GMR.

The magnetic hysteresis loops were determined at room temperature in a vibrating sample magnetometer.

**Results and discussion**

*Composition analysis.* — The overall composition of the multilayers was measured for both series and the results are shown in Fig. 2 as a function of the copper layer thickness. The graph displays the overall concentration of copper in the multilayers. Since the total multilayer thickness was 300 nm, due to the large penetration depth of the electron beam during the analysis (sampling depth is typically 1 μm), the 20 nm thick Cu underlayer also gives a contribution to the measured Cu concentration which is non-negligible especially at low Cu layer thicknesses. The measured data were corrected for this substrate effect in two different ways. First, the energy-dispersive spectra measured with the RÖNTEC analytical facility were evaluated directly for composition without any automatic correction by the RÖNTEC software (i.e., without the so-called ZAF correction). Then, the substrate contribution was subtracted by assuming the Cu and Co contributions to the measured spectra being proportional to the nominal thicknesses (both for the individual layers and for the Cu underlayer). The data obtained in this manner are denoted by the label "corr1" in Fig. 1 (open symbols). In the second correction method, the substrate spectrum measured on a multilayer-free Si/Cr/Cu substrate area was subtracted from the spectra of the multilayer measured together with the substrate. Again, no ZAF correction was applied in evaluating the composition from this substrate-corrected spectra and these data are denoted by the label "corr2" in Fig. 2 (filled symbols). On the average, the two sets of corrected analysis data yield essentially the same trend for both multilayer series.

The chemical analysis results revealed that the Cu content in the G/P series multilayers increases monotonously with Cu layer thickness whereas for the multilayers of the G/P/G series, it is nearly constant (Fig. 2). The observed monotonous increase of the Cu content with $t_{Cu}$ in the G/P series is in accordance with expectation on the basis of the constituent layer thicknesses: the thickness of the Co layer containing about 2.8 at.% Cu only remains constant throughout the whole series whereas the Cu layer thickness increases continuously. According to the analysis results, the composition of the multilayers with the thickest Cu layers agrees very well for the two series. The constantly high Cu-content in the G/P/G multilayers can be explained by taking into account the increased concentration of $Cu^{2+}$ ions at the cathode-electrolyte interface as a result of both the dissolution pulse and the diffusion of $Cu^{2+}$ ions from the bulk solution.

In the G/P/G series, first a 2 nm thick Co layer is deposited with a G pulse and the 6 nm Cu layer is deposited by using a P pulse, then this is followed by the partial dissolution of the previously deposited 6 nm Cu layer with the third, anodic G pulse. After this dissolution process, we apply again a cathodic G pulse for the deposition of a Co layer without any time gap. Due to both the dissolution of the copper layer by the anodic G pulse and the diffusion from the bulk solution, a large excess of copper ions are available in the electrolyte near to the cathode surface and these copper ions also get reduced by consuming a part of the charge applied for cobalt deposition. This leads to an incorporation of an enlarged amount of Cu into the Co layer and this magnetic layer should then contain much more Cu for the G/P/G series than for the G/P series as observed indeed. We can also see from Fig. 2 that for larger Cu layer thicknesses the difference between the overall Cu contents for the two series gradually decreases since less and less Cu is dissolved in the third pulse of the G/P/G sequence.

It should be noted that for the cases of dissolving a large fraction of the 6 nm Cu layer deposited in the P pulse for the G/P/G series multilayers, i.e., for small $t_{Cu}$ values in this series, the 2 nm thick Co layers may also be reached by the dissolution process since Cu dissolution certainly proceeds unevenly over the cathode surface. The partial dissolution of the Co layer finally also leads to an increased overall Cu content in the multilayer, similarly to the excess Cu content at the cathode/electrolyte interface at the end of the Cu dissolution pulse, and the two mechanisms cannot be identified separately from the composition data. However, we experienced that the dissolution potential in the anodic pulse approached the equilibrium potential of the $Cu/Cu^{2+}$ system ($E \approx$ -15 …-60 mV vs. SCE at the end of the dissolution pulse), indicating that the dissolving metal was Cu. In the case of Co dissolution, an abrupt decrease in the dissolution potential is expected (down to about $E \approx$ -300 …-400 mV vs. SCE) when the dissolution front reaches the Co-rich layer at a sufficiently large surface area. Therefore, no evidence can be obtained form the dissolution potential for the dissolution of a significant amount of cobalt. We shall later return again to the possibility of Co dissolution when discussing magnetic and magnetoresistance data.

Even though the total Cu content is larger in the multilayers for the G/P/G series, the bilayer thickness was in accord with that designed for a particular sample since the individual layer thicknesses are controlled by monitoring the charge passing through the cell for each pulse separately.

*Structural study.* — The XRD patterns of all multilayers were measured in the $2\theta$ range from 35° to 55°. For all multilayers, an intense (111) peak and another small peak (200) are observed between the positions of the corresponding Bragg reflections of the face-centered cubic (fcc) phase of pure Co and Cu metals. In Figs. 3 and 4, the XRD patterns are displayed for the G/P and G/P/G series, respectively, around the fcc(111) peaks for a better visibility of eventual satellite peaks due to multilayer periodicity.[14] The XRD patterns were shifted horizontally to the same peak position for each multilayer since this way the evolution of visible satellite reflections with Cu layer thickness (bilayer thickness) can be better observed.

As can be seen in Fig. 3b, the G/P series multilayers exhibit clear satellite reflections $S^-$ and $S^+$ for Cu layer thicknesses of at least 3 nm. The presence of satellites in the G/P multilayers for $t_{Cu} \geq 3$ nm indicates that in these multilayers there is a fairly good coherence of layer growth since this is an important condition for coherent reflections from the subsequent bilayers[14] which finally can yield satellite reflections. The lack of obviously visible satellite reflections for some of the multilayers with thinner Cu layers does not exclude the existence of a definite multilayer structure, it just shows the reduction of the structural coherence along the thickness to a critical degree as a consequence of which the conditions for observing satellite reflections are not fulfilled.

Of the multilayers with Cu layer thicknesses below 3 nm in the G/P series, the samples with $t_{Cu}$ = 1.5 nm, 2 nm and 2.5 nm seem also to have satellite reflections, although mostly very faint ones. The low-angle satellite ($S^-$) for the multilayer with $t_{Cu}$ = 2.0 nm apparently overlaps with the bulk hcp-Co(100) reflection which should appear roughly at the same position as the $S^-$ satellite. Therefore, due to this somewhat sharper peak, we cannot neglect the presence of a hcp-Co(100) reflection here which, then, indicates the occurrence of a very small amount of hcp phase in this particular multilayer.

The XRD pattern of the G/P multilayer with $t_{Cu}$ = 0.5 nm (Fig. 3a) should be discussed separately since it has a peculiar appearance. An enlarged version of this pattern is shown in Fig. 3c together with the result of the XRD line fitting which reveals two additional peaks on the low-angle side of the main peak. By looking at the inserted vertical lines indicating the positions of the XRD lines of the Cu and Co reflections in the angular range displayed, it can be established that the peak at the lowest angle corresponds to a small fraction of hcp-Co phase. This phase often occurs at low spacer thicknesses in ED Co/Cu multilayers[11,15] as it occurred also in the G/P multilayer with $t_{Cu}$ = 2 nm (see Fig. 3a). The origin of the occurrence of such a phase is that the discontinuous nature of the Cu layer at low thicknesses provides a chance for the growth of the Co layer without interruption by fcc-Cu at some locations[11] and this way, after a certain thickness, Co will adopt the stable hcp phase instead of the metastable fcc one. Furthermore, we can see in Fig. 3c that the main fitted peak and the larger minor fitted peak fall within the peak positions of the fcc-Cu(111) and fcc-Co(111) or hcp-Co(002) reflections (the two latter lines are practically at the same position) and they are fairly close to the pure metal line positions. These fitted peaks can, therefore, be identified as corresponding to a pure fcc-Cu phase (minor peak at lower angle) and to a pure fcc-Co or hcp-Co phase (major peak at higher angle). Note that the area ratio of the two peaks is fairly well compatible with the effective layer thickness ratio of the Cu and Co layers (0.5 nm vs. 2.0 nm). This is an indication that in this particular sample, a layered Co/Cu structure could not form due to the discontinuity of the Cu layer since then, as is the case for all the other multilayers, a single main fcc(111) peak ought to have appeared instead of two separate peaks of the two constituent metals. However, since the two phases (Cu and Co) are present in the form of a nanoscale mixture and there is some matching of the lattice planes of the two phases, this leads to a slight shift of the peak positions towards each other. The appearance of obvious separate reflections for the two metals indicates also that their sizes are sufficiently large to relax their respective lattice constants almost to the pure metal values.

At least from the visual inspection of the measured XRD patterns, all the other multilayers are free from the hcp-Co phase and form a layered fcc structure as indicated by the single main fcc(111) diffraction line. The main observed peak can reliably be assigned to an fcc phase since, as was noticed above, and fcc(200) reflection could also be seen for each multilayer and, furthermore, the hcp-Co(101) reflection was completely missing in each measured XRD pattern.

It is noted that the evolution of XRD patterns in the G/P multilayers of the present study, especially concerning the appearance of satellite reflections, are in very good agreement with the XRD results[11] on a previously investigated similar G/P Co(~2.7 nm)/Cu($t_{Cu}$) multilayer series with a total thickness of about 450 nm which was prepared under very similar deposition conditions except for the bath which contained additionally also $H_3BO_3$ and $(NH_4)_2SO_4$.

As to the G/P/G series multilayers, satellite reflections can only be observed for the case of $t_{Cu}$ = 6 nm (Fig. 4b) which was obtained without dissolving away any fraction of the Cu layer deposited in the P pulse (this sample is nominally identical with the last multilayer of the G/P series). The lack of satellite reflections for the G/P/G multilayers with $t_{Cu}$ < 6 nm can certainly be assigned to an increased Cu content in the magnetic layers as was pointed out in our previous work.[16] In such a case, namely, the coherent reproduction of interfaces is diminished and layer thickness fluctuations will occur which, then, lead to a broadening of the satellite peak with a concomitant reduction of its intensity. This can be further amplified by an increased surface roughness to be shown later which is a consequence of the dissolution process.

According to Fig. 3b, for the G/P series with increasing Cu layer thickness, i.e., for increasing bilayer repeat length ($\Lambda = t_{Co} + t_{Cu}$), the satellite peak positions approach toward the main peak as expected.[14] On the basis of the relation given in Ref. 17, from the position of the main and satellite peaks for the (111) reflection which was determined by the fitting procedure as described in the Experimental section, we calculated the bilayer thickness for those G/P multilayer samples where it was possible to establish sufficiently accurately the satellite peak positions. The ratio of the experimental and nominal $\Lambda$ values obtained in the present work is displayed in Fig. 5a by open triangles. The $\Lambda_{exp}/\Lambda_{nom}$ ratio is slightly larger than 1 but it generally corresponds to similar data reported previously for ED Co/Cu multilayers from both XRD studies and direct cross-sectional TEM imaging[11,15,16,18] which are also included in Fig. 5a. The majority of the data is in the range $\Lambda_{exp}/\Lambda_{nom} = 1.1 \pm 0.1$. The much larger values for some of our samples can certainly be attributed to the inaccuracy of the satellite peak position determination due to the low satellite intensity and/or their strong overlap with the main Bragg.

It should be finally noted that for the present G/P and G/P/G series multilayers with $t_{Cu}$ = 6 nm, similar satellites appear and even the bilayer values determined from XRD agree fairly well. The agreement for the two multilayers with the thickest Cu spacer layer comes from the fact that they are nominally identical: they were produced in both series separately and, therefore, these two samples can be considered as a reproduction test. As was the case with the compositional analysis, the reproducibility can be considered as very good also from viewpoint of the multilayer structure, specifically the bilayer thickness.

The FWHM values have also been determined for the main fcc(111) reflection for all the multilayers in both series. An XRD line broadening can occur due to various lattice imperfections defects in the multilayers and larger FWHM values generally indicate stronger structural disorder as discussed in Ref. 11. The FWHM values did not show any systematic evolution with Cu layer thickness for either series and they were ranging between 0.15 and 0.30 deg. Whereas it is not straightforward to derive quantitative parameters for the size and amount of the various structural imperfections, at least these FWHM data are well in conformity with data reported previously for a similar G/P Co/Cu multilayer series with varying Cu layer thicknesses which were prepared from a different bath.[11] In the latter work, FWHM values between 0.27 and 0.30 deg were obtained for multilayers with clear satellite reflections (in the range of $t_{Cu}$ from about 2 to 4 nm). Apparently, the present multilayers have at least the same structural quality since their FWHM values are typically even somewhat lower than the above data from Ref. 11.

We have also evaluated an effective lattice constant from the position of the main fcc(111) reflections (by neglecting the unavoidably present tetragonal distortion of the cubic cells due to the in-plane lattice mismatch of the two constituent layers consisting of different elements). As shown in Fig. 5b, the overall evolution of the lattice parameter roughly corresponds to the results of compositional analysis for the two series which were presented in Fig. 2. For the G/P series, the lattice parameter increases monotonously with $t_{Cu}$ in accord with the change of the overall Cu content. This is actually expected on the basis of the lattice constants of the constituent metals ($a_{Co}$ = 0.35446 nm and $a_{Cu}$ = 0.36148 nm, see Ref. 19). We can also see that these lattice parameter data extrapolate fairly well to the bulk fcc-Co value for $t_{Cu}$ = 0. On the other hand, the lattice parameter data on the G/P/G multilayers show a peculiar behavior in that they exhibit a dip in the middle which is not believed to reflect any real change in the multilayer structure. Instead, we would consider these data as showing an approximately constant behavior, especially if we disregard the data for $t_{Cu}$ = 3 nm in the G/P/G series which may, eventually, be loaded with a larger error than the other data in the series. A definite source of lattice parameter error is that we used the simple approach of getting the lattice parameters from the main Bragg peaks only. The more correct approach is that the XRD patterns are recorded over a broader angular range (in our case, the 2Θ range covered the first two Bragg peaks only) and the lattice parameter is determined from several Bragg peak positions and the data are then extrapolated to Θ = 90º (2Θ = 180º). Nevertheless, even the data presented reveal that for the G/P/G series the lattice parameter is approximately constant with a large scatter and the average value is slightly larger than the G/P multilayer data for the thickest Cu layers. Therefore, disregarding a single lattice parameter data point from the G/P/G series, the lattice parameter data for both series are well in accordance with the composition analysis results (Fig. 2).

*Surface roughness.* — The root-mean-square roughness ($R_q$) values determined through an AFM analysis are presented in Fig. 6. The surface roughness values are rather different for the two series of multilayer samples.

The G/P series multilayers have comparatively smooth surfaces and their root-mean-square surface roughness values varied between 4 to 23 nm with a fairly clear monotonic decrease towards larger Cu layer thicknesses. The higher roughness values for small Cu layer thicknesses may arise from the fact that nucleation of Cu on Co proceeds in an island-like manner[20] and a sufficiently high effective Cu layer thickness should be achieved to ensure a uniform coverage. This mechanism explains clearly the evolution of roughness with $t_{Cu}$ for the G/P series multilayers.

It should be noted, at the same time, that the surface roughness evolution with spacer layer thickness was found to be just the opposite for ED G/P $Ni_{50}Co_{50}$/Cu multilayers.[21] Also, the magnitude of $R_q$ was typically much larger for comparable layer thicknesses and total multilayer thickness in the case of the $Ni_{50}Co_{50}$/Cu multilayers.[21] This difference may eventually come from the presence of additional bath components ($Na_2SO_4$, $H_3BO_3$, $H_2NSO_3$) of the Ni-Co-Cu bath with respect to the pure sulfate bath used in the present work for Co/Cu multilayer deposition. Furthermore, larger $R_q$ values for Ni-Co/Cu multilayers than for Co/Cu and Ni/Cu multilayers were observed also in Ref. 22 when all these multilayers were deposited at the electrochemically optimized Cu deposition potential from the same type of bath as used in Ref. 21. This was attempted to be explained as a consequence of the simultaneous presence of two kinds of magnetic ion in the bath.

In case of the G/P/G Co/Cu multilayer series, the roughness trend was completely different and for most samples in this series, the surfaces were rougher than in the G/P series. It is particularly important to note that by dissolving about 0.5 nm or 1 nm of the previously deposited smooth 6 nm thick Cu layer immediately raises the surface roughness enormously. Upon dissolving a larger fraction of the 6 nm thick Cu layer, the roughness reduces again drastically but it still remains mostly above the values for the G/P multilayers with the same Cu layer thickness, with a slight average decrease of $R_q$ towards low $t_{Cu}$ values. It is interesting to correlate the roughness data with structural parameters derived from XRD. It was shown in the previous section that only G/P multilayers above about 2 nm Cu layer thickness exhibit satellite reflections (and also the G/P/G multilayer with $t_{Cu}$ = 6 nm). By looking at the surface roughness data in Fig. 6, this implies that only multilayers with sufficiently smooth surfaces can give rise to satellite reflections the appearance of which requires the fulfillment of the condition of good structural coherence along the thickness. Evidently, rough surfaces cannot comply with this requirement. This corroborates our recent results on ED $Ni_{50}Co_{50}$/Cu multilayers[21] where it was also observed that satellite reflections appeared for multilayers with similarly smooth surfaces only as in the present G/P Co/Cu multilayers.

*Magnetoresistance data.* — The GMR measurements have been performed for all Co/Cu multilayers and a typical *MR*(*H*) curve is presented in Fig 7. The *MR*(*H*) curves nearly reached saturation in magnetic fields around 2 kOe for most of the samples. The saturation behavior is observed due to the FM regions of the magnetic layers and the non-saturation behavior observed is due to the presence of SPM regions in the magnetic layers. The separation of FM and SPM contributions was done using a standard Langevin fitting process.[5] The result of such a separation process is also demonstrated in Fig. 7. For the particular multilayer chosen, the SPM contribution to the total GMR was fairly small and this was typical for most of the multilayers investigated except for the G/P/G multilayers with the smallest Cu layer thicknesses which will be discussed later separately.

Figure 8a shows the evolution of the total measured GMR in the highest applied magnetic field of 8 kOe as a function of the Cu layer thickness for both the G/P and G/P/G series. In the G/P series, the observed GMR shows the typical monotonous increase[4,21,23-25] with Cu layer thickness from zero GMR until a saturation of about 8 % is achieved at the largest Cu layer thicknesses. On the other hand, the total measured GMR data for the G/P/G series multilayers exhibit much less variation with $t_{Cu}$. For 6 nm Cu layer thickness, the GMR value is somewhat smaller than the corresponding value for the G/P multilayer with the same Cu layer thickness but this small difference is due to the usual uncertainty of the reproducibility of a given sample. For the G/P/G series, on the average, there is a more or less monotonous reduction of the GMR by about a factor of 2 when reaching the smallest Cu layer thickness.

After performing the Langevin-fitting for all the multilayers in order to obtain the saturation values ($GMR_s$) of the FM and SPM contributions to the GMR, the results presented in Fig. 8b were obtained. As seen before for the total GMR, the $GMR_{FM}$ contribution for the G/P series shows the same evolution with Cu layer thickness, just with some smaller values due to the removal of the $GMR_{SPM}$ contribution which is small (typically 1 % or less) with respect to the FM term. For the G/P/G series, the $GMR_{FM}$ contribution shows a decrease by a factor of 3 when going from $t_{Cu}$ = 6 nm to 0.5 nm whereas the $GMR_{SPM}$ contribution increases roughly by the same ratio in this Cu layer thickness change.

In order to understand the different behavior of GMR in the two series, it is instructive to compare the $t_{Cu}$ = 0.5 nm multilayers from the G/P and G/P/G series. It is revealed by Fig. 9 that the *MR*(*H*) curves of these two samples with the thinnest Cu layers show a distinctly different behavior.

As one can see in Fig. 9a, the G/P series multilayer at such a small thickness of the Cu spacer layer exhibits no GMR, only an AMR effect which is a characteristic of bulk ferromagnetic materials[26-28] (LMR > 0, TMR < 0, with the value of the AMR defined as the difference LMR - TMR). Such a behavior is typical for ED Ni-Co/Cu multilayers with very small spacer thickness for any ratio of Ni to Co in the magnetic layers when prepared with the usual two-pulse plating (G/P or P/P) (Refs. 4,15,20,21,23-25). The origin of this bulk-like behavior lies in the presence of a large density of pinholes in the Cu layer[4,6,15,24] which gives rise to a direct FM coupling between adjacent magnetic layers.

On the other hand, the measured *MR*(*H*) curve of the G/P/G multilayer also with $t_{Cu}$ = 0.5 nm exhibits a distinctly different character since even the LMR component is negative for the whole range of magnetic fields investigated as revealed by Fig. 9b (the TMR component not shown was very similar, except for a slightly larger magnitude), i.e., this multilayer exhibits GMR. The decomposition of the measured magnetoresistance in FM and SPM contributions is also given in Fig. 9b and we can establish that in this particular sample the SPM contribution is roughly of the same magnitude as the FM contribution. It can be inferred from Fig. 8b that the $GMR_{SPM}$ contribution for the G/P/G series multilayers increases gradually when the Cu layer thickness reduces down to below about 3 nm by the dissolution process. This Cu layer thickness range roughly corresponds to the thickness range where the overall multilayer Cu content data indicated a larger Cu content in the magnetic layer with respect to the G/P series multilayers. Therefore, the primary reason for the increase of the relative importance of the $GMR_{SPM}$ contribution on the account of the $GMR_{FM}$ contribution lies in the fact that due to the increased Cu-content in the magnetic layers, a phase separation takes place which gives rise to the appearance of Co segregations in the form of SPM regions as it was shown for ED Co/Cu multilayers by both MR measurements[16,25] and direct structural studies.[16]

As discussed at the end of the subsection on the results of compositional analysis of the present multilayers, it was noticed that for small $t_{Cu}$ values in the G/P/G series, the 2 nm thick

Co layers may also be reached by the dissolution process. This can eventually also lead to a partial fragmentation of the Co layers, providing another pathway for the formation of SPM entities in the magnetic layer. Nevertheless, the MR data are either not appropriate to decide whether the excess $Cu^{2+}$ ion concentration or the Co dissolution leads finally to the observed increased Cu content in the G/P/G multilayers.

Finally, it is noted that the above interpretation of the difference in the magnetoresistance results between the G/P and G/P/G multilayers with $t_{Cu} = 0.5$ nm are fully in agreement with the structural studies. Namely, in the G/P multilayer which exhibited AMR, the XRD investigation indicated the presence of a dominant Co phase and a separate minor Cu phase. The ratio of the two phases is in conformity with a picture of a percolating magnetic (Co) phase which is expected to yield a bulk-like AMR behavior. In the G/P/G multilayer, the observed GMR necessitates the presence of a fairly well-defined layered structure. Although satellite reflections were not observed in this sample, the XRD study still revealed that a single main fcc(111) peak occurs which corresponds to an average common lattice plane distance of the Co and Cu layers as expected for a nanoscale layered structure of two metals.[14]

*Magnetic properties.* — In order to characterize the magnetic behavior of the multilayers, the hysteresis loops were measured for each sample up to a magnetic field of 12.5 kOe. The hysteresis loops became closed in typically a magnetic field of about 2 kOe and this ensured that a complete magnetic saturation of the ferromagnetic regions could be achieved in these fields. This is supported by the magnetoresistance results shown in Figs. 7 and 9b where we can observe that the decomposed $GMR_{FM}$ contribution in both cases indeed reached saturation at around 2 kOe.

The evolution of the coercive field $H_c$ with Cu layer thickness is shown in Fig. 10a for both multilayer series. The behavior of $H_c$ for the G/P series is qualitatively the same as reported in our previous work[4] for an ED G/P Co/Cu series with a magnetic layer thickness of about 2.7 nm. The coercive field is low for small Cu layer thicknesses where mainly AMR dominates the observed magnetoresistance due to the pinholes in the thin Cu layers. With increasing Cu layer thickness, the decoupling of the magnetic layers from each other by the Cu spacer layers becomes more and more efficient and this results in an increase of the coercive field to a saturation value characteristic for the given thickness of an individual Co layer. This behavior of the coercive field also supports the explanation of the gradual increase of the $GMR_{FM}$ component for the G/P series (see Fig. 8b) with Cu layer thickness as being a consequence of the more random alignment of the adjacent layer magnetizations in zero field due to the reduction of the pinhole-induced FM coupling in comparison with multilayers having smaller Cu layer thicknesses.

We can also observe in Fig. 10a that the magnitude of $H_c$ for the present G/P series is larger than previous data reported in Ref. 4. This is partly connected with the fact that in Ref. 4, the magnetic layer thickness was about 2.7 nm whereas in the present case it was 2 nm since thinner magnetic layers are known to have higher coercive fields. Apart from the magnetic layer thicknesses, also the electrodeposition baths were different for the two G/P series (this work and Ref. 4) which might have resulted in different microstructural features and/or internal stresses. This could also be an explanation for the observed different coercivities although, as discussed above, the FWHM linewidths sensitive to the microstructure and stresses were of comparable magnitude for the two series.

The overall evolution and magnitude of the coercive field of the G/P/G series matched fairly well that of the G/P series. For large Cu layer thicknesses, this is not so surprising since also the $GMR_{FM}$ data are very similar here (see Fig. 8b). The upturn of the $H_c$ data for the G/P/G series for small Cu layer thicknesses can, on the other hand, have some significance. According to Fig. 8b, the large $GMR_{SPM}$ component here indicates the presence of an increased amount of SPM regions but since for these the coercive field is zero, they do not contribute to the observed coercive field. Therefore, the observed upturn of $H_c$ can be ascribed to the increased coercive field of the FM regions at low Cu layer thicknesses in the G/P/G multilayers. An increase of the coercive field can be expected if the size of the FM regions is reduced. Certainly this is the case since the large amount of SPM regions here suggests that the magnetic part of the sample is definitely split up into smaller regions which exhibit either FM or SPM characteristics. A smaller size of the FM region may be a reduction either in its thickness or its lateral extension; both features lead to an increased coercive force.

The behavior of the reduced remanence of the multilayers is also interesting (Fig. 10b). For the G/P series, the relative remanence is fairly high (around 0.9) although with a slight decrease towards larger Cu layer thicknesses. The large remanence is in conformity with the small $GMR_{FM}$ contribution since it is an indication of the absence of an antiferromagnetic interlayer coupling; the slight decrease of the remanence towards larger Cu layer thicknesses hints at a more random alignment of adjacent layer magnetizations in zero field, yielding then an increase of the $GMR_{FM}$ term. For the G/P/G series, on the other hand, we can observe a strong reduction of the remanence with decreasing Cu layer thickness. This is in agreement with the increasing amount SPM regions indicated by the decomposed MR measurements (Fig. 8b). At the lowest Cu layer thickness of the G/P/G series, half of the observed GMR is arises due to spin-dependent scattering events along electron pathways between a FM and a SPM region so the volume fraction of the SPM region should be non-negligible here anymore. As a consequence, we should expect a reduction of the overall remanence as actually observed.

## Summary

In this work, ED Co(2 nm)/Cu($t_{Cu}$) multilayers with $t_{Cu}$ ranging from 0.5 nm to 6 nm were fabricated by using two different deposition pulse combinations after the basic electrochemical parameters had been optimized through cyclic voltammetric and chronoamperometric techniques. The first series was prepared by using a G/P pulse combination whereby the galvanostatic (G) pulse was used to deposit a magnetic Co layer and the potentiostatic (P) pulse was applied for a non-magnetic Cu layer deposition. In the second series, a G/P/G pulse combination was used to prepare first a Co(2 nm)/Cu(6 nm) bilayer in each cycle as in the case of the last sample of the G/P series and, then, the Cu layer was gradually dissolved by the anodic third G pulse to achieve a preset Cu layer thickness. The purpose of the work was to carry out a comparative study of the composition, structure, surface roughness and GMR on the two series in which a given Cu layer thickness was achieved in two different ways of multilayer preparation.

Significant differences were observed between the two series in the various parameters investigated. A chemical analysis revealed that whereas for the G/P series the overall multilayer Cu content varies with $t_{Cu}$ as expected on the basis of layer thicknesses, the G/P/G multilayers for low $t_{Cu}$ values exhibit an increased Cu content which can mainly be assigned to the incorporation of excess Cu in the magnetic layers. This latter feature could be explained as arising due to the increased Cu ion concentration at the cathode/electrolyte interface at the start of the Co deposition pulse as a consequence of the preceding anodic G pulse applied for Cu dissolution.

The comparison of the results of an XRD study and a surface roughness analysis by AFM has shown that only those multilayers exhibit multilayer satellite reflections for which the surface roughness is sufficiently small, in agreement with our recent results on ED $Ni_{50}Co_{50}$/Cu multilayers.[22] G/P multilayers with $t_{Cu}$ below about 2 nm and all the G/P/G multilayers with Cu dissolution (i.e., all but multilayer with $t_{Cu}$ = 6 nm) had larger surface roughness values that prevented the occurrence of a coherent reflection from the subsequent bilayers and this can well explain the absence of satellite reflections for these multilayers. The dissolution of the first 0.5 nm and 1 nm of the 6-nm thick Cu layer in the G/P/G series caused an enormous roughness increase which then again strongly reduced by further Cu layer dissolution although the surface roughness remained always higher than that observed for the corresponding G/P multilayer.

An analysis of the magnetoresistance data revealed that the G/P series multilayers, apart from the smallest Cu layer thickness (0.5 nm), exhibit GMR typically with a relatively small SPM contribution only. The magnitude of the GMR increases monotonously with $t_{Cu}$ and finally reaches saturation as found also in previous studies of ED multilayers.[4,21,23-25] The total GMR of the G/P/G multilayers showed much less variation with Cu layer thickness but the magnetoresistance decomposition analysis has shown that the $GMR_{FM}$ contribution reduces strongly towards smaller Cu layer thicknesses. At the same time, the relative weight of the SPM contribution increased here and this could mainly be ascribed to an increased Cu content in the magnetic layers for small Cu layer thicknesses.

When comparing the G/P and G/P/G multilayer for $t_{Cu}$ = 0.5 nm, an absence of GMR and the presence of an AMR effect was observed for the G/P case whereas a definite GMR with a comparable magnitude of the FM and SPM contributions could be obtained for the G/P/G case. This clearly indicates that the Cu layers at such small thicknesses are indeed different depending on whether they are formed in a single deposition step (G/P series) or depositing a thick Cu layer first and then partially dissolving it (G/P/G series). The direct deposition of thin Cu layers (G/P sequence) proceeds via island formation with pinholes remaining between the islands and then a coalescence of Cu islands occurs as the effective Cu layer thickness increases. When the thin Cu layer is formed by dissolving a thick Cu layer (G/P/G sequence), evidence was found from the GMR results that FM regions in adjacent magnetic layers can occur which are not coupled ferromagnetically to each other. A comparison of the XRD patterns of these two particular multilayers strongly supported this picture.

The magnetic hysteresis loops yielded coercive field and remanence data for the same multilayers. An analysis of these data could be carried out in the same picture as outlined above for the magnetoresistance and thus provide strong support for the interpretation put forward for the evolution of microstructure with Cu layer thickness in both series and also for explaining the observed differences between the two series.

The difference in the properties of two-pulse and three-pulse plated samples of nominally identical layer structure clearly reveals a significant difference in the microstructure of the two sample groups. In analyzing the observed differences, we should keep in mind, however, that the dissolution and the deposition processes cannot be regarded as merely the opposite of each other. Therefore, the entire electrode process has to be scrutinized as a whole, and low- and high-current pulses cannot be automatically associated with the deposition of non-magnetic and magnetic metals, respectively. Besides the sample composition, the surface morphology also exhibits a significant difference as a function of the sample preparation procedure. This may also open up new ways in the field of modulated deposition methods for reaching otherwise inaccessible sample structures and compositions.

The present study demonstrated that the layer formation can be effectively controlled via various pulse combinations although, evidently, further steps are necessary to avoid the unwanted consequences occurring due to the increased $Cu^{2+}$ ion concentration caused by the partial dissolution of the spacer layer.

**Acknowledgements** We are indebted to the Hungarian Scholarship Board for providing a 7-month fellowship to N.R. for his research stay in Budapest. The authors also acknowledge G. Molnár (Institute for Technical Physics and Materials Science, Research Centre for Natural Sciences, HAS) for preparing the evaporated underlayers on the Si substrates. This work was supported by the Hungarian Scientific Research Fund (OTKA) through Grant K 104696.

**Figures**

Fig. 1 (a) Cyclic voltammetry curve for the Co-Cu electrolyte used. The inset shows an enlargement of the potential range for Cu deposition; (b) Current transient curves for various copper deposition potentials.

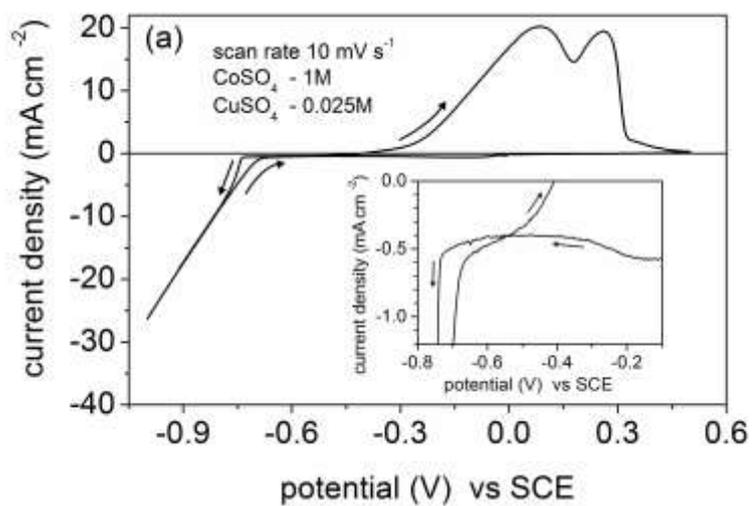

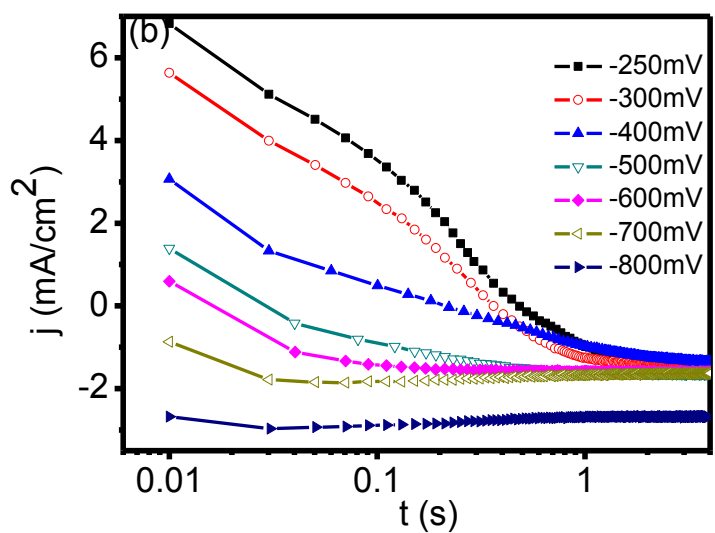

Fig. 2 Results of the composition analysis for the G/P and G/P/G series multilayers. The overall Cu content in the multilayer is shown as a function of the nominal Cu layer thickness $t_{Cu}$ which was derived from the parameters of the applied pulses. The measured composition data were corrected for the substrate Cu underlayer contribution by two methods (corr1 and corr2) as explained in the text.

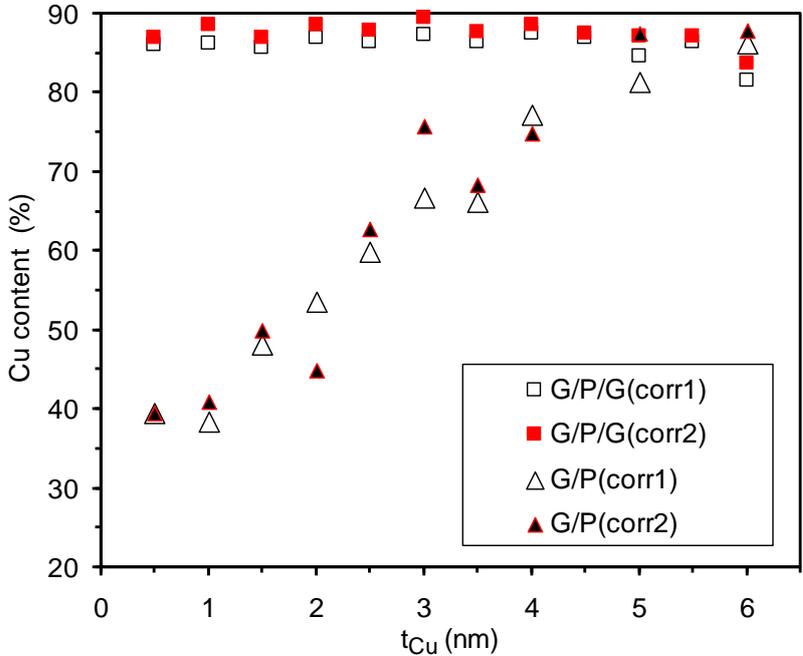

Fig. 3 X-ray diffraction patterns of the G/P series multilayers with various copper layer thicknesses ($t_{Cu}$) as indicated in the range (a) 0.5 nm to 2.5 nm and (b) 3.0 nm to 6.0 nm. Please note that the fcc (111) peak positions of all multilayers were shifted horizontally to the same position in order to better visualize the evolution of satellite peak positions ($S^+$ and $S^-$) with copper layer thickness. (c) enlarged view of the XRD pattern for $t_{Cu}$ = 0.5 nm with the results of fitting.

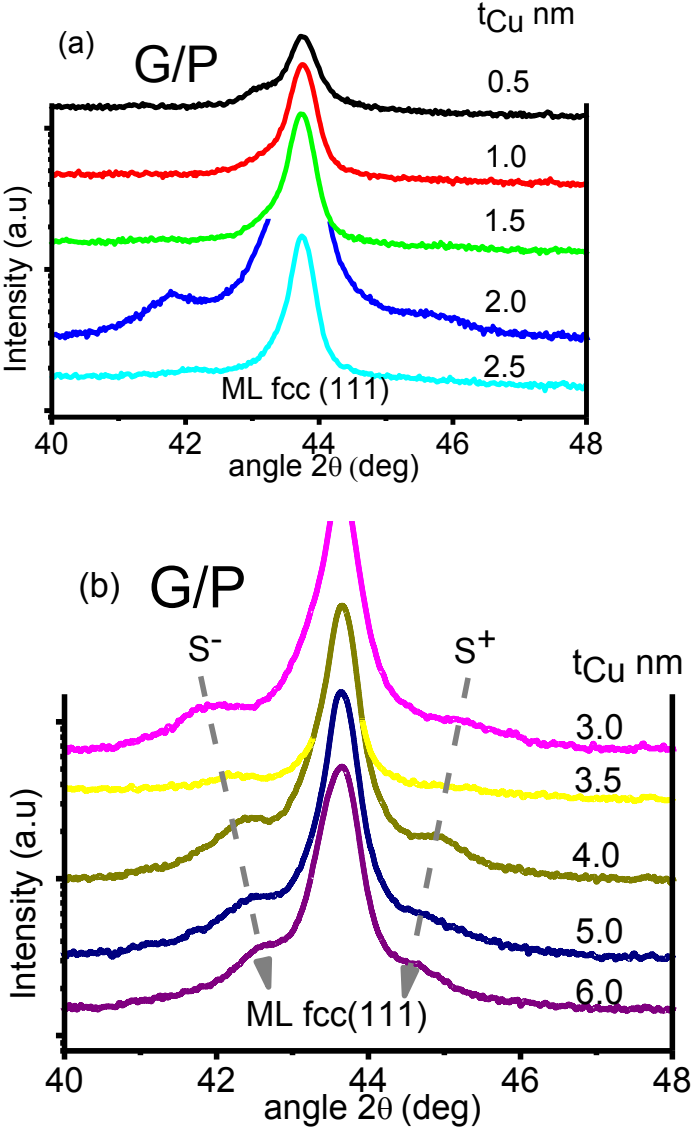

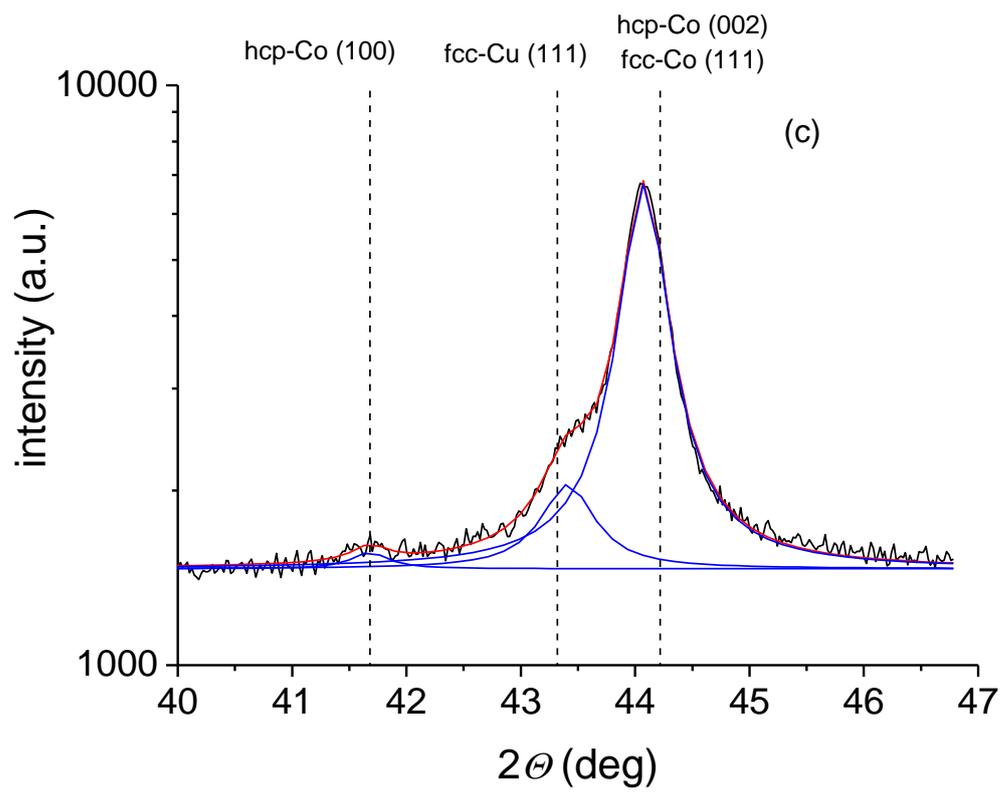

Fig. 4 X-ray diffraction patterns of the G/P/G series multilayers with various nominal Cu layer thicknesses ($t_{Cu}$) as indicated in the range (a) 0.5 nm to 3.0 nm and (b) 3.5 nm to 6.0 nm. Please note that the fcc (111) peak positions of all multilayers were shifted horizontally to the same position in order to better visualize the evolution of satellite peak positions ($S^+$ and $S^-$) with copper layer thickness.

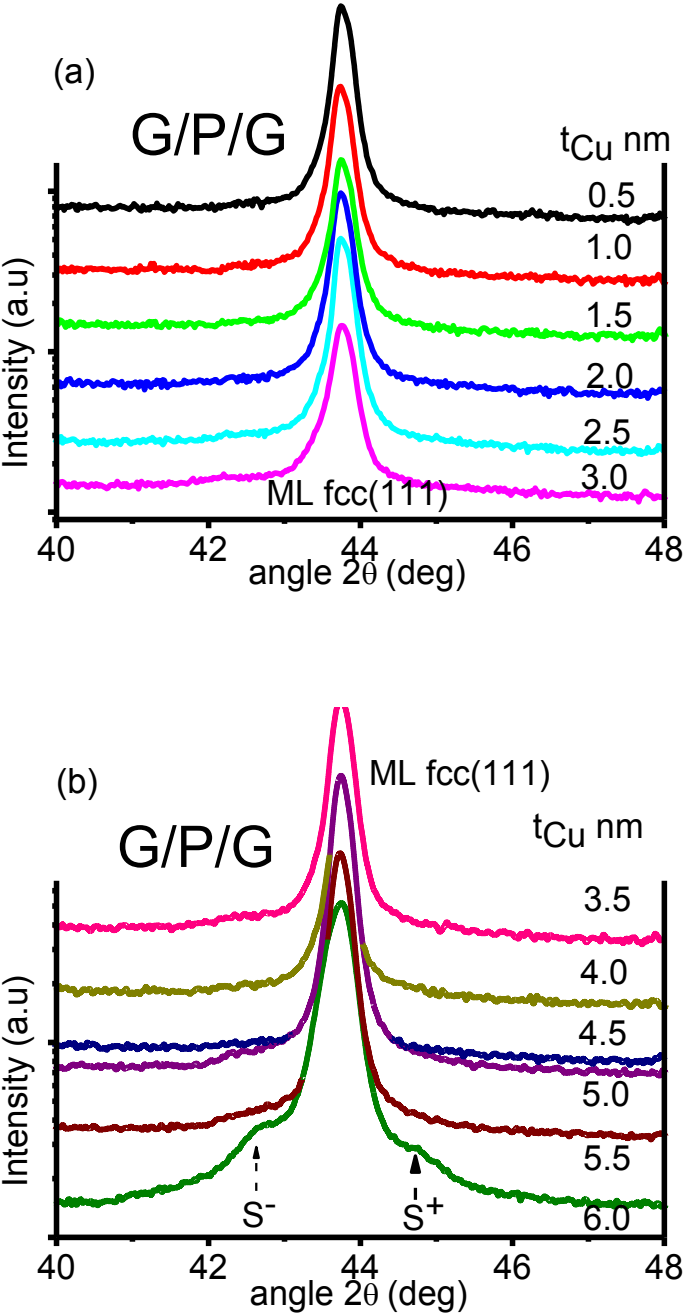

Fig. 5 (a) Bilayer repeat length $\Lambda_{exp}$ as determined from the XRD satellite peak positions and normalized with the nominal bilayer repeat length $\Lambda_{nom} = t_{Co} + t_{Cu}$ for the G/P series multilayers. Data from previous XRD and TEM studies on ED Co/Cu multilayers are also included as indicated in the legend; (b) Lattice constant '$a$' for the G/P and G/P/G series multilayers as evaluated from the fcc(111) XRD line positions.

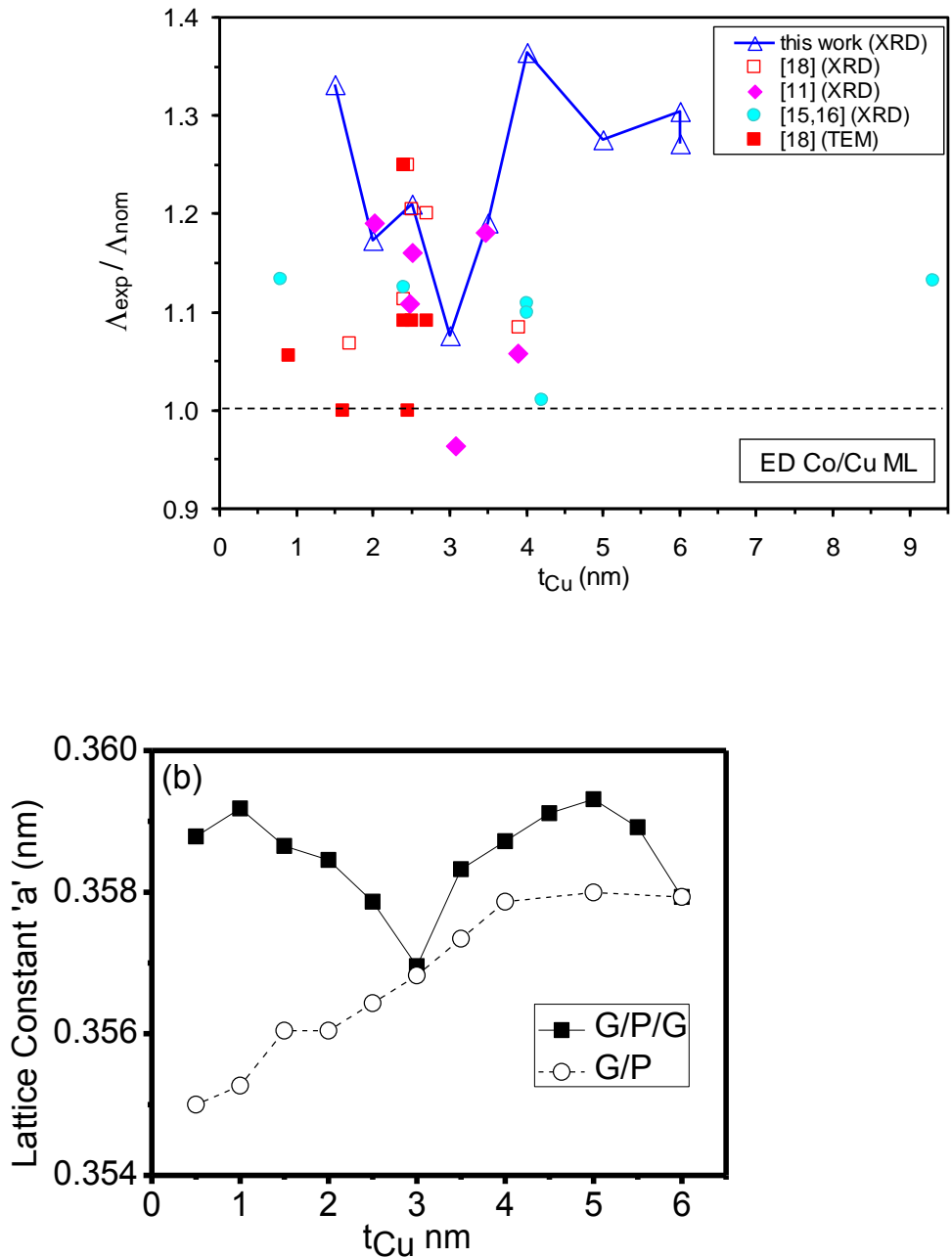

Fig. 6 Root-mean-square surface roughness ($R_q$) of the G/P and G/P/G series multilayers derived from AFM analysis as a function of the Cu layer thickness ($t_{Cu}$).

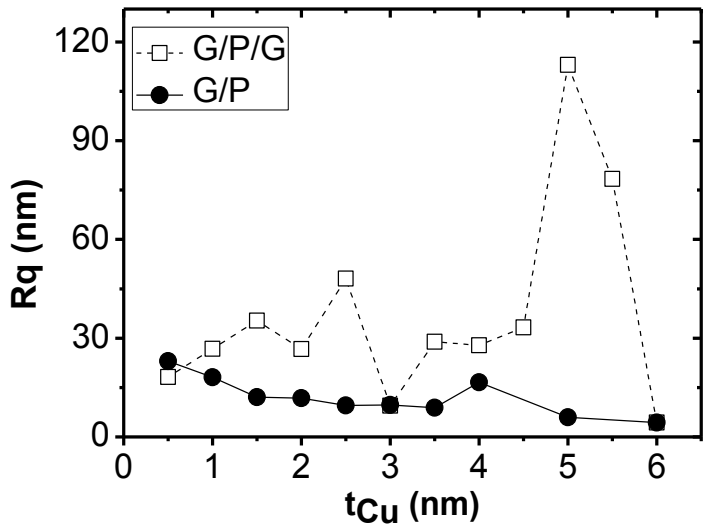

Fig. 7 A typical measured $MR(H)$ curve (TMR component) for the [Co(2 nm)/Cu(3.5 nm)]×55 multilayer from the G/P/G series with the results of the Langevin fitting yielding the $GMR_{FM}$ and $GMR_{SPM}$ contributions as indicated. The shape of the $MR(H)$ curve for the LMR component was the same, just with a smaller magnitude.

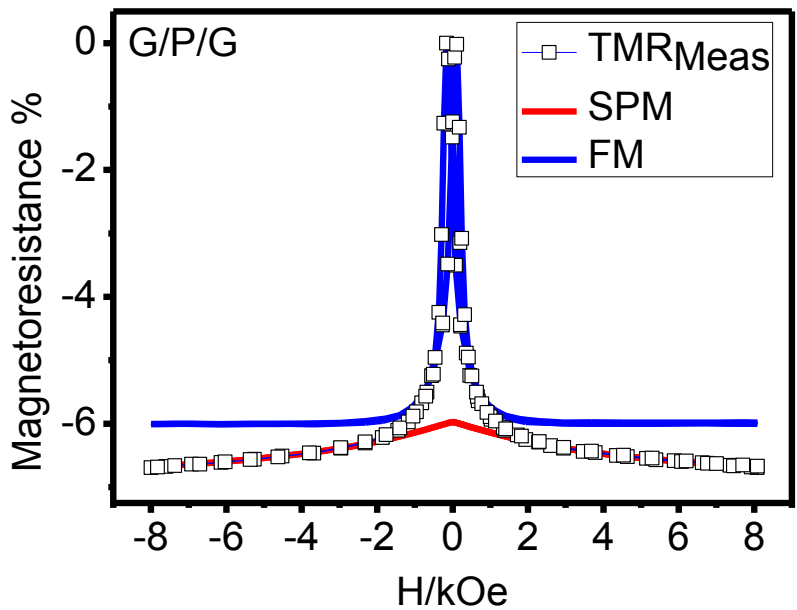

Fig. 8 *(a)* The total GMR measured at the maximum applied field (8 kOe) for both the G/P and the G/P/G series multilayers as a function of the Cu layer thickness ($t_{Cu}$); *(b)* Decomposed saturation $GMR_{FM}$ and $GMR_{SPM}$ contributions for both G/P and G/P/G series multilayers as a function of the Cu layer thickness. Since there was a slight difference only between the LMR and TMR values, in order to avoid confusion due to the too much data points, in (b) the average of the LMR and TMR values is displayed only.

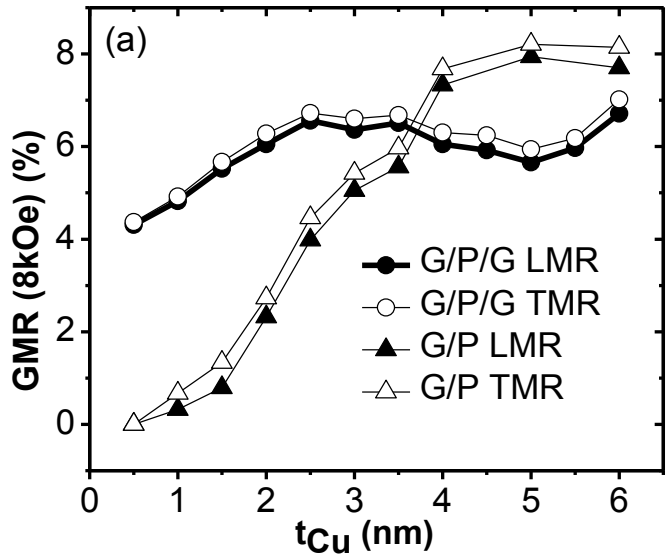

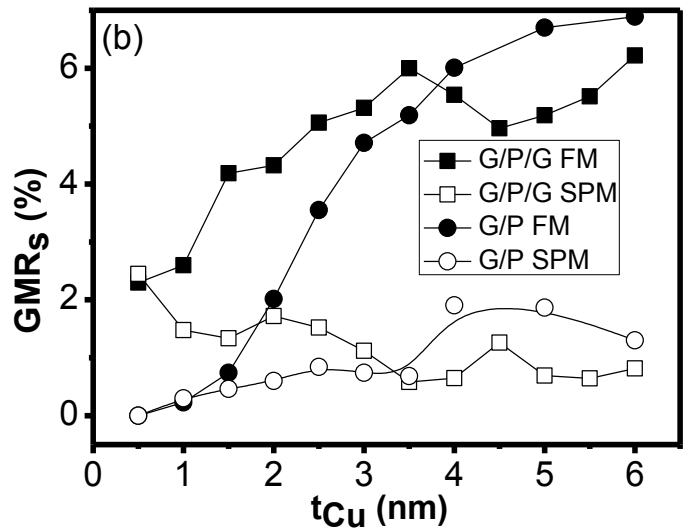

Fig. 9 *(a)* Measured longitudinal and transverse *MR(H)* curves for the G/P multilayer [Co(2 nm)/Cu(0.5 nm)]×120 which exhibits AMR (LMR > 0; TMR < 0); *(b)* Measured longitudinal *MR(H)* curve for the G/P/G multilayer [Co(2 nm)/Cu(0.5 nm)]×120 by showing also the results of GMR decomposition into FM and SPM components.

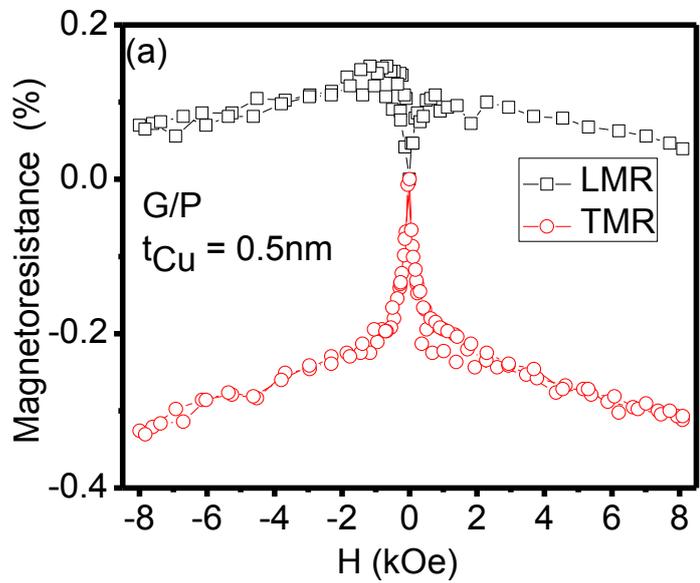

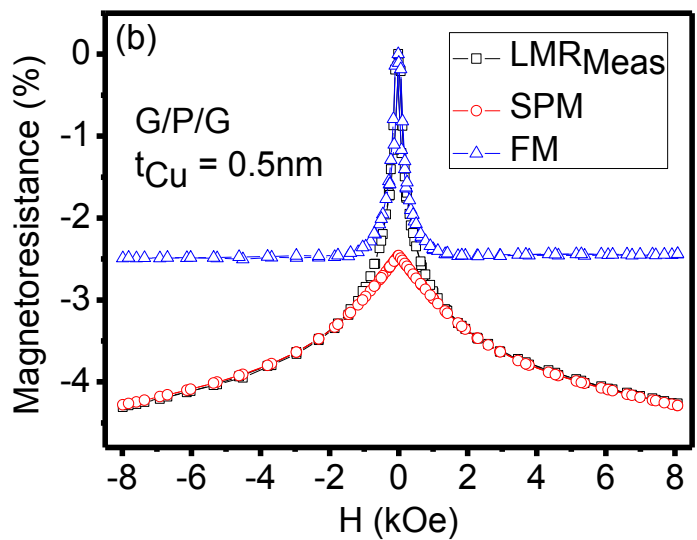

Fig. 10 Evolution of the coercive field $H_c$ *(a)* and the relative remanence $M_r/M_s$ *(b)* for both the G/P and the G/P/G series multilayers with nominal Cu layer thickness ($t_{Cu}$).

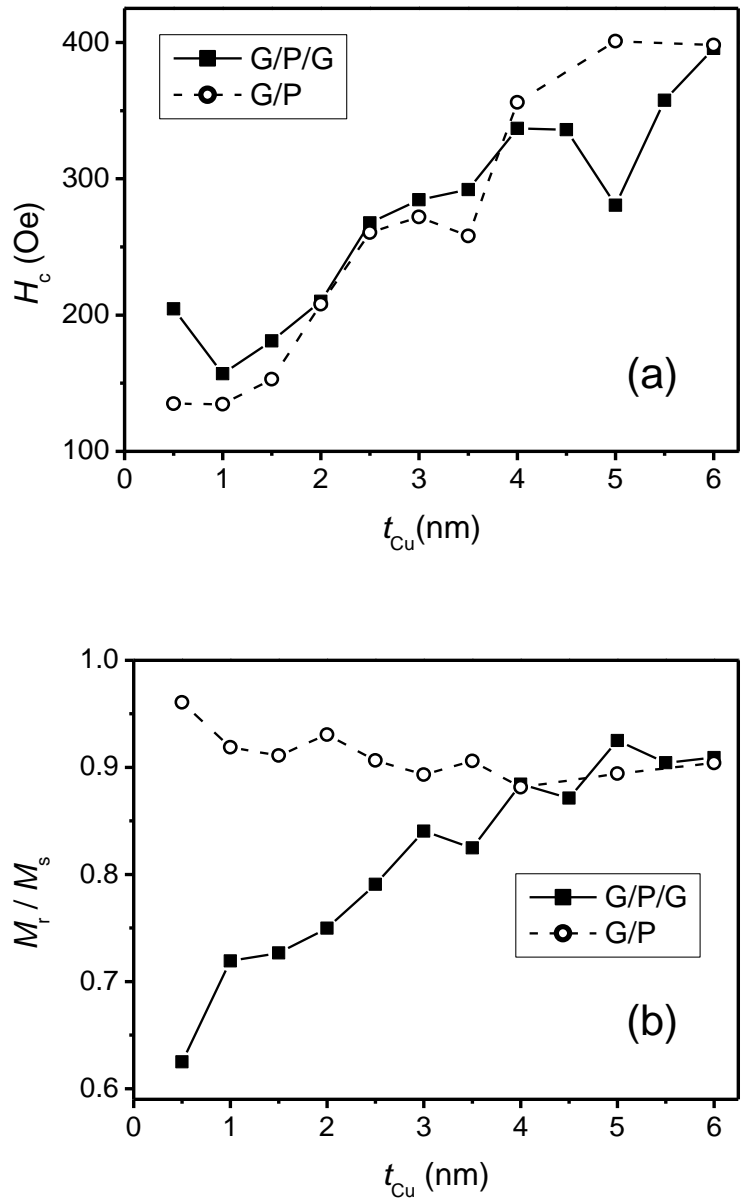